\documentstyle[prb,aps,preprint]{revtex}

\begin{document}


\title{Novel aspects and strong correlation in the electronic structure of
Sr$_2$FeMoO$_6$}

\author{Sugata Ray$^1$, Priya Mahadevan$^{1, 2}$, Ashwani Kumar$^1$, D. D. Sarma$^{1,}$ \cite{jnc}}

\address{$^1$Solid State and Structural Chemistry Unit, Indian Institute
of Science, Bangalore 560~012, India \\}

\address{$^2$National Renewable Energy Laboratory, Golden, Colorado 80401\\}

\author{R. Cimino}

\address{LNF-INFN V. E. Fermi 40 I-00044 Frascati (Roma), Italy
\\}

\author{M Pedio, L. Ferrari}

\address{ISM-CNR sede distaccata Trieste S.S. 14,
Km. 163.5 I-34012 Basovizza (TS), Italy\\}

\author{A. Pesci}

\address{University of Trieste, Trieste, Italy\\}

\maketitle

\begin{abstract}
We investigate the electronic structure of Sr$_2$FeMoO$_6$
combining photoemission spectroscopy with a wide range of photon
energies and electronic structure calculations based on 
first-principle as well as model Hamiltonian approaches 
to reveal several interesting aspects. We find
evidence for unusually strong Coulomb correlation effects both in
the Fe 3$d$ {\em and} O 2$p$ states, with an enhanced
manifestation in the majority spin channel. Additionally, O 2$p$
states exhibit a spin-splitting of nonmagnetic origin, which
nevertheless is likely to have subtle influence on the stability
of novel ferromagnetism of this compound.

\end{abstract}

PACS number(s): 79.60.-i, 71.20.-b, 71.27.+a, 71.20.Ps


\vspace*{1cm}

\newpage
\vspace{0.5cm}
\begin{center}
{\bf Introduction}
\end{center}
\vspace{0.5cm}

Sr$_2$FeMoO$_6$ has been recently shown to exhibit a pronounced
negative magnetoresistance at higher temperatures and lower
magnetic fields \cite{koba} compared to the manganites, making it
intrinsically more suitable for technological applications. An
essential aspect of understanding the properties of this material
is to understand the underlying electronic structure. The basic
ingredient for crystal structure of this material is the cubic
perovskite structure, with Fe and Mo alternating in the (100),
(010) and (001) directions. As in the case of the other $AB$O$_3$
perovskites, where $A$ is an alkali-metal atom and $B$ is a transition
metal atom, the valence and conduction bands of Sr$_2$FeMoO$_6$
involve the $d$ states of the two transition metal atoms - Fe and
Mo, besides the oxygen $p$ states. It has been recently
shown~\cite{Mn1,T1,COSSMS} that the Mo~4$d$-O~2$p$ hybridized
states play a crucial role in establishing the unusually high
magnetic ordering temperature in Sr$_2$FeMoO$_6$ {\it via} a new
mechanism that relies on the location of these Mo~4$d$-O~2$p$
hybridized states energetically between the exchange-split Fe
states. There has been only one direct investigation \cite{Saitoh}
of the electronic structure of this compound so far, with another
investigation \cite{Min} reporting on a related compound,
Ba$_2$FeMoO$_6$. These experimental results suggest substantial
disagreement with {\it ab initio} band structure results and are
interpreted in terms of LDA+U calculations by incorporating
Coulomb interactions within the Fe 3$d$ manifold. We have
performed photoemission experiments on Sr$_2$FeMoO$_6$ and
Sr$_2$FeMo$_x$W$_{1-x}$O$_6$ with a photon energy range that
encompasses the combined range of the previous two experiments,
farther extended by including photoemission results with x-rays.
These detailed experimental results are analyzed in conjunction
with calculations based on multiband Hubbard model within
mean-field decoupling of the interaction terms as well as {\it ab
initio} approaches. 

We identify the origin of various features in the experimental
valence band spectrum with the help of the dependence of the
matrix elements on the photon energy. Thus, we find that while the
specific ordering of energy levels near the Fermi energy, crucial
for the magnetism, is in agreement with the {\it ab initio}
results, there are some significant differences between the
experimental and the {\it ab initio} results in agreement with
previous reports, indicating the presence of strong
correlation effects. Moreover, we show that the conventional LDA+U
and parametrized model Hamiltonian approaches are also in
disagreement with the experimental results in certain details,
particularly in the spectral range dominated by oxygen $p$ states,
though the inclusion of $U_{dd}$ improves the description of
Fe~$d$ related spectral features as also observed previously. It
turns out that the inclusion of the interaction terms within the
oxygen $p$ states improves the agreement between the calculated
results and the experiment. We further establish a few unusual
aspects of the electronic structure in this compound specifically
related to the O $p$ band, such as a spin-splitting of nonmagnetic
origin. These findings are possibly of importance in understanding
the stability of the ferromagnetic state and provide a clue why
the magnetic structure in this compound has been recently found to
depend critically on factors that influence the O $p$ states.
\cite{T2,H1}

\vspace{0.5cm}
\begin{center}
{\bf Experimental}
\end{center}
\vspace{0.5cm}

Polycrystalline Sr$_2$FeMoO$_6$ was prepared by standard solid
state route as described in ref.~1. Stoichiometric amounts of
SrCO$_3$, MoO$_3$ and Fe$_2$O$_3$ were mixed thoroughly and heated
at 900$^{\circ}$C in air for 3 hrs. The fully oxidized black
powder is then grounded and pelletized into 10 mm. diameter
pellets of 2 mm. thickness. These pellets were then annealed at
1200$^{\circ}$C in a flowing gas mixture of $\sim$10\% H$_2$/Ar
for 5 hrs. X-ray diffraction experiments on this powder confirmed
the presence of highly ordered ({\it i.e.} near perfect 
alternate occupancy of Fe and Mo
ions along the three cubic axes) and pure double perovskite
Sr$_2$FeMoO$_6$ phase. It has tetragonal I4/mmm space group with 
$a$ = $b$ = 10.5229 a.u. and $c$ = 14.9236 a.u.; 
Sr, Fe, Mo, O1 and O2 sites
occupy 8g, 4e, 4e, 4e and 8h positions, respectively.

A different method was applied for synthesizing 
W-doped Sr$_2$FeMoO$_6$ samples, because it was otherwise not possible
to entirely 
get rid of SrMoO$_4$/SrWO$_4$ type oxidized impurities, whenever the
W-doped samples were synthesized by the above mentioned procedure. Therefore, 
all the W-doped samples were first prepared by melt-quenching
method~\cite{ssc};
the technique which always produced slightly less ordered but highly pure
samples. These pellets were crushed, grounded thoroughly and
finally annealed at the same condition in the form of pellets
in order to achieve maximum ordering.~\cite{jpcm} The
purity and chemical homogeneity of the final products were checked
by XRD and Energy Dispersive Analysis of X-rays (EDAX) techniques.

Majority of the photoemission experiments were carried out at the
VUV photoemission beamline, 3.2R at Synchrotron Radiation Centre,
Elettra, Trieste, with a sample temperature of 77 K. The sample
surface was cleaned by {\it in situ} scraping with a diamond file.
XPS valence band spectrum was recorded in a standard VSW
electron spectrometer \cite{krishna} at $\sim$120~K with AlK$\alpha$
radiation. 

\vspace{0.5cm}
\begin{center}
{\bf Calculational Methods}
\end{center}
\vspace{0.5cm}

The {\it ab-initio} band structure of Sr$_2$FeMoO$_6$ was computed 
using both the LMTO-ASA method using the Perdew Wang GGA exchange 
functional \cite{pw91} and the plane wave pseudopotential method. 
No empty spheres were introduced, and the volume filling criterion was satisfied with
less than 16\% overlap between the spheres in the LMTO-ASA calculations. 
Ultrasoft pseudopotentials \cite{usp} as implemented in VASP \cite{vasp}
were used for the calculations performed with a plane wave cutoff of 20~Ry and the accuracy 
of our conclusions was checked by increasing the plane wave cutoff to 25~Ry. 
Both the internal as well as the external cell parameters were optimized.
The GGA \cite{pw91} optimized lattice constants were 10.59 and 14.997 a.u. 
and the optimized internal coordinates were Sr (8g) z=0.25, Fe (4e) z=0, Mo (4e) z=0.5, O1 (4e)
z=0.2527 and O2 (8h) x=0.2474, in agreement with the experimental parameters.
 
The band structure obtained from both {\it ab-initio} methods were found to be 
in agreement. We fit the magnetic band structure 
along various symmetry directions with band dispersions within a nearest neighbor 
tight binding model \cite{JPhysC} in order to obtain estimates of the hopping
parameters for the use in the multiband
Hubbard model. The basis of the tight-binding Hamiltonian 
includes $d$ orbitals on Fe and Mo atoms as well as $p$ orbitals on oxygen.
As the Sr states do not contribute in the energy window that we are interested 
in, the basis of the tight binding model does not include 
states on the Sr atoms.
The tight binding model considered includes hopping between 
oxygen atoms, as well as between Fe and oxygen and Mo and oxygen; these interaction 
strengths \cite{params} are given in terms of the Slater Koster parametrization. 
Instead of using different onsite energies ($\epsilon_i$) 
for up and down spin orbitals to describe the magnetic structure as in the
past,~\cite{JPhysC} we replace the different on-site energies for the 
up and down spin of Fe 
by an on-site direct Coulomb ($U_{dd}$) interaction strength of
3 eV and an exchange Hund's coupling ($J$) strength of 0.7 eV on the Fe
site as appropriate~\cite{JPhysC} for Fe$^{3+}$ oxide systems; this method allows us to 
solve for the changes in the energies of the Fe 3$d$ level as a result of direct 
and exchange Coulomb interactions
self-consistently. Besides the usual one-body tight binding part,
the Hamiltonian includes the intra-site Coulomb interaction term at the Fe site 
in the form  
$(U_{dd}$-$J$$\delta_{\sigma_1,\sigma_2}$)$n_{\alpha_1, \sigma_1}$ $n_{\alpha_2, \sigma_2}$,
where $\alpha_i$ denotes the five 3$d$ orbitals on the Fe site and $\sigma_i$ denotes the 
spin. As discussed later in the text, certain features in the experimental spectra
prompted us to include an additional term representating 
Coulomb interactions within the oxygen $p$ orbitals in the form 
$U_{pp}$~$n_{\beta_1, \sigma}$ $n_{\beta_2, \sigma^{'}}$. Effect of such a term involving
$U_{pp}$ has been studied in the context of the electronic structure of the
of high temperature superconductors~\cite{ref2} and other oxide systems.~\cite{anisimov}
The four fermion terms in the Hamiltonian were decoupled by using a 
mean-field approximation.~\cite{hf} The decoupled Hamiltonian involved the occupancies
of various orbitals, which were then solved selfconsistently 
over a k-mesh of 1000 k-points.

\vspace{0.5cm}
\begin{center}
{\bf Results and discussion}
\end{center}
\vspace{0.5cm}

In Fig.~1 we show a selection of the valence band photoemission spectra
of Sr$_2$FeMoO$_6$, obtained using different incident photon
energies. All the spectra exhibit a finite intensity at $E_F$, in
agreement with the metallic behavior in this compound.
For comparison, we also show the calculated total density of
states (DOS) for fully ordered Sr$_2$FeMoO$_6$, obtained within
spin-polarized LMTO-ASA band structure calculation with
generalized gradient approximation (GGA) over the relevant energy
range just below the experimental spectra. The experimental
spectra show four distinct features, appearing at about 0.4, 1.8,
5.6 and 8.3~eV binding energies, marked by the vertical dotted lines in Fig.~1.
The calculated LMTO-DOS clearly does not agree with the measured
spectra, with the calculation suggesting regions of low DOS at three of these
energies, namely at 1.8, 5.6 and 8.3 eV. In particular, the
remarkably intense feature at 8.3~eV in the experimental spectra
lies outside the calculated bandwidth. This is similar to the
cases of U-intermetallics~\cite{U} and transition metal oxides,
such as NiO~\cite{nio}, exhibiting correlation-driven large
spectral weights, or even distinct peaks, outside the calculated
bandwidth. It is significant to note that this correlation driven
satellite in Sr$_2$FeMoO$_6$ is very intense and is undoubtedly
the highest among all Fe$^{3+}$ oxides studied so
far~\cite{ashish,dd}; for example, it is known \cite{dd} that
spectroscopic data on LaFeO$_3$, a closely related perovskite
oxide with Fe$^{3+}$, can be reasonably well described within {\it
ab initio} approaches.

A standard route to take into account the effect of Coulomb
interactions on the band structure is to incorporate this
interaction within the transition metal 3$d$ manifold in a
mean-field sense, as in the case of LDA+U approach or the  
Hartree-Fock (HF) type mean-field approach described for
the multiband Hubbard model in the previous section. It is indeed
true that such approaches have proved to be very useful for
describing transition metal compounds. We
show the total density of states obtained within the HF-treatment of
the multiband Hamiltonian including Coulomb interactions only at the Fe sites at
the bottom of Fig. 1; our result is in agreement with LDA+U
results \cite{T1,Saitoh,Min,H1}. A comparison of the experimental
data with the two calculations (LMTO and HF)readily suggests that the HF-type
approach with the multiband model provides a better agreement with
the experiment in comparison to the LMTO-GGA results.
Specifically, the experimentally observed 1.8 eV feature and the
satellite structure at about 8 eV, outside the GGA derived
bandwidth, have corresponding features in the model Hamiltonian
results. Interestingly, however, the sharp disagreement between
the most intense feature in the 5-6~eV range in the experimental 
spectrum and the deep valley of
low calculated DOS in the same energy range continues to exist in
the HF-DOS, just as in the LMTO-DOS. The same
discrepency can also be observed in the LDA+U results compared
with experimental results, reported in ref. [5, 6, 8] though this
disagreement has not been commented up on so far.

Before attempting to understand the origin of this discrepency, we first
establish the origin of various experimental features from the
dependencies of the normalized spectral intensities on the photon
energies (Figs. 2(a) and (b)). We found that the two features
appearing at 1.8 and 8.3 eV exhibit significant non-monotonic
$h\nu$ dependence in the range of 42-65 eV (Fig. 2(a)), spanning the
Fe 3$p$ absorption threshold near 55 eV. Such a resonance
phenomenon clearly establishes substantial Fe 3$d$ contributions
in these states. More interestingly, the nature of the
dependencies is different for the two features. The 8.3 eV feature
exhibits a clear maximum, indicating about 30\% resonant
enhancement, characteristic of such correlation driven satellite
or incoherent features. We have found almost identical variations
in the satellite features in closely related W-substituted
compounds also, as shown in the figure. In contrast, the 1.8 eV
feature shows approximately a 20\% dip in the intensity near the
on-resonant condition, as is characteristic of more band-like or
coherent features.~\cite{fe}

In order to identify the states with dominant Mo contributions, we
plot the normalized intensities of the features at the $E_F$ 
and 8.3 eV for $h\nu$ between 75 and 152~eV in Fig.~2(b). This photon energy
range is most suitable for the purpose, since there is a minimum
(called Cooper minimum) in the cross-section of localized Mo 4$d$
related intensities in solids at about $h\nu \sim$ 100-110 eV.~\cite{abbati}
Fig. 2(b) clearly shows a significant minimum in
the relative intensity of the 8.3~eV feature (triangles) at about
$h\nu \sim$ 100 eV, exactly where the Cooper minimum for the
localized Mo 4$d$ states are expected. This establishes that the
Mo 4$d$ states also contribute significantly into the 8~eV
spectral region. In contrast to the behavior of the localized
states, delocalized band-like Mo $d$ states, however, do not
exhibit any Cooper minimum.~\cite{abbati} Therefore, the relative
intensity of Mo 4$d$ delocalized states compared to the O 2$p$
states is expected to show a rapid, monotonic increase with
increasing photon energy, $h\nu$, due to the changes in the
relative Mo 4$d$ : O 2$p$ cross-sections.~\cite{yeh} Such an
increasing trend is indeed exhibited by the relative intensity of
the feature at $E_F$ (see square data in Fig. 2(b)), establishing a
dominant contribution from {\em delocalized} Mo 4$d$ states in
this energy region.

The assignment of various spectral features are further confirmed
by a comparison of the intensity variations when the photon energy
is changed from $h\nu$~$<$~100~eV to 1486.6 eV using AlK$\alpha$
radiation. In Fig. 3 we show these spectra with the 
integrated area being normalized for each spectrum, with two
representative low photon energy spectra with $h\nu$~=~54.6~eV
(open circles) and $h\nu$~=~86.4~eV (open squares), while the
x-ray photoelectron spectrum (XPS, shown with open triangles in
Fig. 3) with $h\nu$~=~1486.6~eV. It is evident from this
comparison that there are substantial changes in the spectral 
intensities of various features, as the photon energy is changed by 
$\sim$~1400~eV, thereby
changing the relative cross-sections of various states
drastically. The most obvious spectral changes are the very strong
increases in the intensities of features near {\it E$_F$} and in
the 7-9~eV range, substantial increase in the intensity of the
feature at 1.8~eV and a decrease in the energy region 4-6~eV
binding energy in the high photon energy XP spectrum. Noting that
the Mo 4$d$ cross-section increases strongly and that of Fe 3$d$
moderately relative to the oxygen 2$p$ states, we may attribute
substantial Mo 4$d$ contributions in the near {\it E$_F$} and
7-9~eV binding energy region, primary Fe 3$d$ contribution in the
1.8~eV region and a dominance of O 2$p$ derived states in the
4-6~eV binding energy. These conclusions are in agreement with the 
interpretation based on the spectral changes within the low photon
energy ($h\nu$~$<$~200~eV) range (see Figs. 2(a) and (b)). Fig. 2
also established substantial localized Fe 3$d$ contributions in the
neighborhood of the 8.3~eV feature, in addition to the Mo 4$d$
contribution. These assignments
of the spectral features at once provide us with an understanding
of both the partial success and the specific failure of the mean-field
calculated results (HF-DOS) as well as LDA+U results reported
earlier~\cite{T1,Saitoh,Min,H1} in the literature, to 
explain the experimental
spectral features. Compared to GGA, the incorporation of Coulomb
interaction within the Fe 3$d$ manifold in terms of these
mean-field approaches improves the agreement between the
experiment and the calculation for the 1.8 and 8.3 eV features
that have significant Fe 3$d$ character (see Fig. 1); this feature
also attains substantial Mo 4$d$ characters due to large
hopping interactions present in the system. However, at
this level of mean-field approximation, there is still a rather
pronounced disagreement between the experiment and calculation in
the 4-6~eV energy range;~\cite{note} our experimental results have already 
established that this energy region is dominated by O $p$ states. 
Thus, a single-particle approach for the O $p$ states fails to describe 
the experiment, while the 
avreage effect of Coulomb interactions within Fe 3$d$ states
improve the description of the Fe 3$d$ related features; this
suggests a possible role of Coulomb interactions also within the 
O $p$ states. This is not altogether surprising, since it is known
from analysis of the {\it KLL} Auger spectra for O in transition 
metal oxides~\cite{ref1,ddold} that the Coulomb interaction strength, $U_{pp}$, in the
O $p$ states is comparable to the Coulomb interaction strength in
the transition metal 3$d$ states. While $U_{pp}$ is not often 
explicitly used in various approaches to describe the electronic
structure of transition metal oxides, same as in the LDA+U method applied 
so far to Sr$_2$FeMoO$_6$ or the preceeding mean-field results
presented in the manuscript, there are some applications of
models including the $U_{pp}$ term~\cite{ref2,anisimov}, most notably in the case of
copper compounds with substantial $p$ admixed character of
the charge carriers. Since the charge carriers
in Sr$_2$FeMoO$_6$ have significantly admixed Mo 4$d$-O 2$p$
character~\cite{COSSMS}, it is reasonable to anticipate a possible
role of $U_{pp}$ also in the present system, thus suggesting an
explanation for the disagreement between the
experimental spectra and the calculated results with $U_{dd}$ alone.
Therefore, we included the $U_{pp}$ term in the multiband Hubbard
model and calculated the total and partial
densities of states within the mean-field
approximation; we found that the inclusion of $U_{pp}$~=~3.8 eV improves the
agreement between the experiment and the calculation substantially,
primarily {\it via} a renormalization of oxygen related site
energies. The agreement improves farther with an enhancement of
the splitting of O $p$ up- and down-spin bare energies 
from the value (0.4~eV) estimated from {\it ab initio} band 
structure results, to 0.7 eV;
the resulting Fe $d$, Mo $d$ and O $p$ partial DOS are
compared with experimental spectra in Fig. 3. Specifically, the
dip in the HF-DOS with $U_{pp}$ = 0 as well as in GGA 
results (Fig. 1) in the energy range of 4-6 eV is replaced with a large
intensity feature in presence of $U_{pp}$, in agreement with the
experimentally observed broad maximum. We also find that the
partial contributions from Mo 4$d$, Fe 3$d$ and O 2$p$ states in
different energy regions are in good agreement with experimental
results. Thus, the calculation shows dominant Mo 4$d$ contribution
due to $t_{2g}$ states at $E_F$, with significant contributions 
 of $t_{2g}$ and $e_g$ symmetries also in the energy range
of 7-9 eV, dominant Fe 3$d$ contributions at 1.5 and 3 eV binding
energy regions, arising from $e_g$ and $t_{2g}$ states respectively,
with a significant intensity contribution in the satellite
region of 8-9 eV and a dominant O 2$p$ nonbonding contribution in
the 4-6 eV binding energy.

Spin-polarized DOS along with the total DOS, shown in Fig. 4, establish 
some interesting features in the electronic structure of this compound.
First, the spectral feature appearing near 8.3 eV, outside the GGA
calculated bandwidth and associated with correlation-driven
satellite, is entirely contributed by the up-spin channel,
suggesting a more pronounced localized nature of electron states
in this spin channel compared to the down spin channel. This
result can be understood in terms of the
half-metallic nature of the ground state with the up-spin channel
indeed localized with a gap at $E_F$, while the down-spin channel
is delocalized with a finite DOS at $E_F$. However, a gap at the
$E_F$ does not necessarily imply a localization of the
wave-function in space; we confirm the localized-in-space nature
of the up-spin channel in contrast to the down-spin channel by
plotting the charge densities of the $e_{g\uparrow}$ and
$t_{2g\downarrow}$ spin states close to $E_F$ in the basal plane
of the unit cell in Fig. 5. This figure shows the extended
nature of the down spin charge density with substantial weights on
all atomic sites, namely Fe, O and Mo. In clear contrast, the
up-spin charge density has virtually no contribution on Mo;
consequently, the charge density plot suggests almost completely
localized charge densities decoupled and isolated on individual
FeO$_6$ octahedra.

We have already pointed out that an analysis of the {\it ab-initio}
band structure results suggests a splitting of about 0.4 eV between
the oxygen up- and down-spin $p$ states and that a proper simulation
of the experimental spectra required a further enhancement of this
splitting  to 0.7 eV within the HF approach. This is an unusually
large spin-splitting compared to oxygen $p$ states in other
ferromagnetic oxide systems. One explanation for the
spin splitting could be based on the atomic exchange strength.
However, since the exchange splitting is given by the product of
the moment and the exchange strength and the typical oxygen moment in this
compound is less than 0.1 $\mu_B$~\cite{tanu}, this explanation would require an
unphysically large exchange strength at the oxygen sites. We have
traced back the non-magnetic origin of this spin-splitting to the
differing spatial extents of the corresponding oxygen charge
densities in up and down spin channels. The overall behavior of 
the charge densities in the two spin channels are grossly similar
and a direct comparison of the two will not show any remarkable
difference. In order to enhance the contrasting behaviors between the 
two spin-channels, we construct the charge density difference between
the two spin channels. In Fig.~6 we have plotted
this difference in up- and down-spin charge densities in the basal
plane. We show the positive part of the $(up~-~down)$ charge density
in the left panel and the complementary $(down~-~up)$ one in the
right panel. Thus, the left panel shows the space region with a
higher charge density in the up-spin channel compared to the
down-spin channel and the right one shows the excess charge density in 
the down-spin channel compared to the up-spin channel. It is clear 
from the figure that the down-spin channel
of oxygen $p$ states are spatially more extended than the up-spin
states, with the spatial extension more dominantly in the
direction of perpendicular bisector of the Fe-O-Mo bond through
the O site. We have analyzed several states within the oxygen
non-bonding region; all of them show the common features of more
extended down-spin channel with relatively more charge density
away from the nearby positively charged metal centers compared to
the up-spin channel. This
shows that the crystal-field stabilization of the electron states
due to the positively charged neighboring sites will be relatively
more for the up-spin oxygen states compared to the down-spin ones,
as indeed found here, giving rise to the apparent exchange splitting 
of the oxygen $p$ states.

In conclusion, we have found that the electronic
structure of Sr$_2$FeMoO$_6$ is dominated by strong
electron correlation effects, manifesting more prominently within
the Fe 3$d$ up-spin states, as a consequence of its half-metallic
ground state. Detailed analysis of the spectral features using
photon energy variation allows us to establish two interesting
effects related to the oxygen $p$ states. First, it appears
necessary to explicitly account for the Coulomb interactions
within the oxygen $p$ states and second, there is a splitting
between the oxygen $p$ up- and down-spin channels arising from
different spatial extensions of the corresponding wave-functions.
In this context, we note that the ferromagnetic stability in
Sr$_2$FeMoO$_6$ is sensitively dependent on the properties of
oxygen states and the first principle band structure calculations
tend to underestimate the ferromagnetic stability.~\cite{T2,H1}
Our findings here are likely to provide an understanding for the stability
of ferromagnetism in this system, as our preliminary calculation of
the spin wave dispersions indicates that the ferromagnetism is
indeed favored by both finite $U_{pp}$ and the ``spin"-splitting of
the oxygen $p$ states.

{\bf Acknowledgement} : We thank O. K. Anderson and O. Jepsen for providing the
LMTO
code. This project is supported by the Department of Science and Technology,
Government
of India.

\begin{figure}
\caption{Valence band spectra of Sr$_2$FeMoO$_6$ with
different $h\nu$ compared with the calculated DOS using LMTO-ASA and 
Hartree-Fock with finite $U_{dd}$.} 
\end{figure}

\begin{figure}
\caption{Percentage variation in the relative intensities of various spectral 
features in the valence band spectra as a function of photon energy 
(a) between 45 and 65~eV for Sr$_2$FeMo$_x$W$_{1-x}$O$_6$;
and parent Sr$_2$FeMoO$_6$
and (b) between 70 and 152~eV for Sr$_2$FeMoO$_6$. All intensities are normalized 
with the intensity of the most intense peak at 5.6 eV, almost fully 
contributed by O $p$ states. The normalized intensity of each feature 
recorded with the lowest $h\nu$ is always scaled to 100\% in order to emphasize
the variation in the relative intensity of various features with $h\nu$}
\end{figure}

\begin{figure}
\caption{Normalized valence band spectra of Sr$_2$FeMoO$_6$ with
$h\nu$ = 54.6, 86.4 and 1486.6 eV; spin integrated partial O 2$p$~(solid line), 
Mo 4$d$~(dashed line), and Fe 3$d$~(short dashed line) DOS obtained by 
Hatree-Fock calculations including $U_{pp}$ in addition to $U_{dd}$ are shown at the 
bottom. Different spectral features are marked by vertical thin lines.}
\end{figure}

\begin{figure}
\caption{Total DOS obtained by Hatree-Fock calculations including 
$U_{pp}$ in addition to $U_{dd}$ (thick solid line). The up (thin solid line)
and down (dashed line) spin DOS are also shown at the bottom of the panel.}
\end{figure}

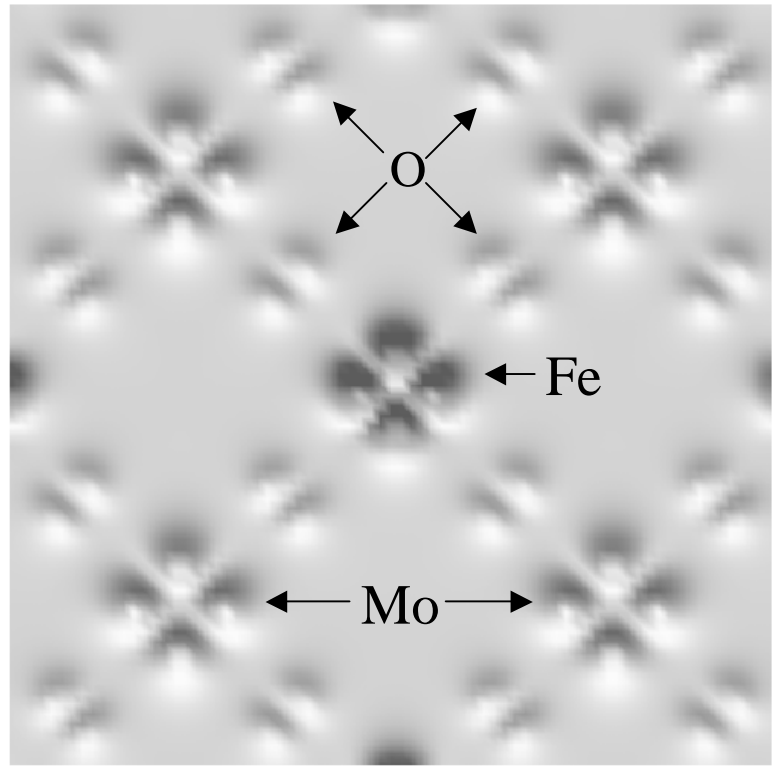
\begin{figure}
\caption{Charge density plots in the basal plane for (a) $e_{g\uparrow}$ states
in the energy window $E_F$-2 eV to $E_F$ and (b)
$t_{2g\downarrow}$ states in the energy window $E_F$-1
to $E_F$+0.5 eV.}

\end{figure}

\begin{figure}
\caption{Plots of differences in charge densities ((a) up - down; and
(b) down - up) of the oxygen non-bonding states at
the $\Gamma$ point in the basal plane.}
\end{figure}


\begin{references}
\bibitem{note}This disagreement is less visible in a comparison between XPS
spectrum and HF results due to a suppression of photoemission cross-section of
O$p$ states at higher photon energies.
\end{references}
\end{document}